\documentstyle[twocolumn,pra,aps,epsfig]{revtex}
\def\be{\begin{equation}}
\def\e#1{\label{#1}\end{equation}}
\def\bea{\begin{eqnarray}}
\def\ea#1{\label{#1}\end{eqnarray}}
\def\r#1{(\ref{#1})}
\def\bem#1{\begin{mathletters}\label{#1}}
\def\eml{\end{mathletters}}
\def\df{\Delta\varphi}

\begin{document}
\draft
\title{Frequent observations accelerate decay: The anti-Zeno effect}
\author{A. G. Kofman and G. Kurizki}
\address{
Department of Chemical Physics, Weizmann Institute of Science,
76100 Rehovot, Israel}

\maketitle
\begin{abstract}
The quantum Zeno effect (QZE) is the striking prediction that the 
decay of {\em any} unstable quantum state can be inhibited by 
sufficiently frequent observations (measurements).
The consensus opinion has upheld the QZE as a {\em general} feature 
of quantum mechanics, which should lead to the inhibition of 
{\em any} decay. 
The claim of QZE generality hinges on the assumption that successive 
observations can in principle be made at time intervals too short for 
the system to change appreciably. 
However, this assumption and the generality of the QZE have scarcely 
been investigated thus far.
We have addressed these issues by showing that 
(i) the QZE is principally unattainable in radiative or radioactive 
decay, because the required measurement rates would cause the system 
to disintegrate; 
(ii) decay {\em acceleration} by frequent measurements (the anti-Zeno
effect -- AZE) is much more ubiquitous than its inhibition.
The AZE is shown to be observable as the enhancement of tunneling
rates (e.g., for atoms trapped in ramped-up potentials or in 
current-swept Josephson junctions), fluorescence rates (e.g., for 
Rydberg atoms perturbed by noisy optical fields) and photon 
depolarization rates (in randomly modulated Pockels cells).
\end{abstract}
\pacs{{\em Keywords}: quantum decay, quantum measurements, Zeno 
effect, anti-Zeno effect, quantum tunneling.}

\section{Introduction}

According to a prevailing view, claimed to be a general feature of
quantum mechanics, successive frequent measurements 
{\em must slow down the decay of any unstable} system 
\cite{kha68,fon73,mis77,sak94,joo84,coo88,ita90,kni90,fre91,pan96,sch98,fac98,ela00}.
This is known as the Quantum Zeno Effect (QZE), introduced by
Misra and Sudarshan \cite{mis77}, following the early work of Khalfin 
\cite{kha68} and Fonda \cite{fon73}.
It has been colloquially phrased as follows: ``a watched arrow never 
flies", or "a watched pot never boils" \cite{kni90}.
Recent estimates of the time interval between measurements required
for the QZE inhibition of radiative decay have been 
$\tau_{\rm Z}\sim 3.6\times 10^{-15}$ s \cite{fac98}.
But is this view true?

\begin{figure}[htb]
\centerline{\epsfig{file=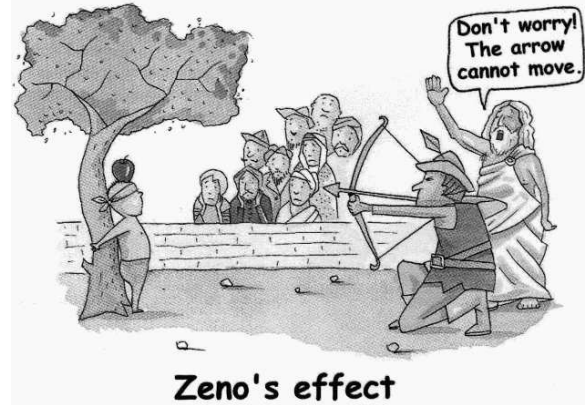,width=3.in}}
\vspace{.3cm}
\centerline{\epsfig{file=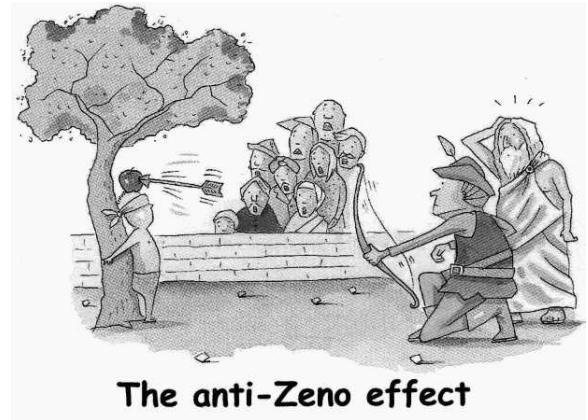,width=3.in}}
\vspace{.3cm}
\protect\caption{
The Zeno and anti-Zeno effects as viewed by Zeno and William Tell
watching an arrow in mid-air.
}
\label{f1}
\end{figure}

We have recently shown \cite{kof00} that, in fact, the 
{\em opposite} is mostly true for decay into open-space continua:
The anti-Zeno effect (AZE), i.e. decay acceleration by frequent
measurements, is far more ubiquitous than the QZE.
This {\em universal} conclusion \cite{lan83,kof96,lew00} has been 
described by 
our commentators in the words: ``a watched pot {\em boils faster}'' 
\cite{mil00}, ``furtive glances trigger decay'' \cite{sei00}.
Accordingly, our cartoonist shows that old Zeno, trying to reassure
Gullielmo Tell and his boy that the arrow won't fly, is in for a nasty
surprise (Fig. \ref{f1}).
How can this conclusion be understood and what was missing in standard
treatments that claimed the QZE universality?

\section{General analysis}

\begin{figure}[htb]
\centerline{\epsfig{file=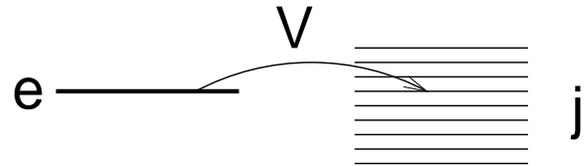,width=3.in}}
\protect\caption{
The decay of a state into a ``reservoir'' via coupling.
}
\label{f2}
\end{figure}

We have reached the conclusion stated above by general analysis of 
measurement-affected decay.
Consider $|e\rangle$, the measured state in a system ruled by
hamiltonian $H=H_0+V$, where $V$ causes the coupling (decay) of 
$|e\rangle$ to all other eigenstates of $H_0$, to which we refer as
the ``reservoir'' (Fig. \ref{f2}).

The probability amplitude $\alpha(t)$ to remain in $|e\rangle$, which 
has the energy $\hbar\omega_a$ obeys the following {\em exact} 
integro-differential equation 
\be
\dot{\alpha}=-\int_0^tdt' e^{i\omega_a(t-t')}\Phi(t-t')
\alpha(t').
\e{1.10}
Here $\alpha(t)=\langle e|\Psi(t)\rangle e^{i\omega_at}$, 
$\hbar\omega_a$ is the energy of $|e\rangle$, and
\be
\Phi(t)=\hbar^{-2}\langle e|Ve^{-iH_0t/\hbar}V|e\rangle=
\hbar^{-2}\sum_j|V_{ej}|^2e^{-i\omega_jt}
\e{1.11}
is the reservoir correlation function, expressed by 
$V_{ej}=\langle e|V|j\rangle$, where $|j\rangle$ ($\ne|e\rangle$) are
$H_0$ eigenvectors with eigenvalues $\hbar\omega_j$.

Equation \r{1.10} is {\em exactly} soluble, but it is enough to 
consider its short-time behavior by setting 
$\alpha(t)\approx\alpha(0)=1$ in the integral of (1).
This yields the expression
\be
\alpha(t)=1-\int_0^tdt'(t-t')\Phi(t')e^{i\omega_at'},
\e{1.13}
in which {\em all powers of} $t$ (phase factors!) are included and
interferences between various decay channels may occur.
By contrast, the standard {\em quadratic} expansion in $t$ for the 
population \cite{mis77,sak94} 
$\rho_{ee}(t)=|\alpha(t)|^2\approx 1-t^2/\tau_Z^2$, in which the Zeno 
time 
$\tau_Z=\hbar/(\langle e|H^2|e\rangle-\langle e|H|e\rangle^2)^{1/2}$
is the inverse variance of the energy in $|e\rangle$, may
often fail, as discussed below.
This is where we essentially differ from standard treatments.
How does this difference show up?

Consider instantaneous measurements -- projections on $|e\rangle$
interrupting its decay at intervals $\tau$.
We can use our result for $\alpha(t)$, Eq. \r{1.13} to express 
the population of $|e\rangle$ after $n$ such measurements as 
exponentially decaying at a rate $R$, 
\be
\rho_{ee}(t=n\tau)=|\alpha(\tau)|^{2n}\approx\exp(-Rt).
\e{33}
The {\em universal form} of $R$ is (in the frequency domain)
\be
R=2\pi\int_0^\infty d\omega G(\omega)F(\omega).
\e{1.16}
This expression is the overlap of the reservoir-coupling spectrum
\be
G(\omega)=\frac{1}{\pi}\mbox{Re}\int_0^\infty dt\Phi(t)e^{i\omega t}=
\hbar^{-2}\sum_j|V_{ej}|^2\delta(\omega-\omega_j)
\e{1.18}
and the measurement-induced broadening of the measured energy level
\be
F(\omega)=\frac{\tau}{2\pi}\mbox{sinc}^2\left(
\frac{(\omega-\omega_a)\tau}{2}\right).
\e{1.19}

\begin{figure}[htb]
\centerline{\epsfig{file=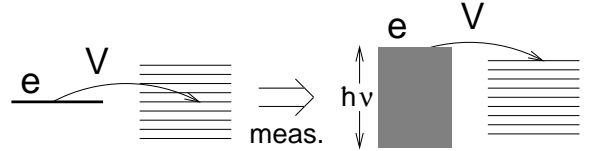,width=3in}}
\vspace{.5cm}
\protect\caption{
Measurements broaden level $|e\rangle$, analogously to 
phase randomization by collisions at rate $\nu$, being drastically
changing its decay into the reservoir.
}
\label{f3}
\end{figure}

We can interpret the universal result (4) as an expression of
the {\em time-energy uncertainty relation} for an unstable level with 
lifetime $\Delta t$, relating the energy broadening (uncertainty) of 
$|e\rangle$ to $\Delta t$, the interval between measurements
(Fig. \ref{f3}),
\be
\mbox{Eq. \r{1.16}}\quad\Longrightarrow\quad\Delta E\Delta t\sim\hbar.
\e{1.20}
More generally, in Eq. \r{1.20} $\Delta t=1/\nu$, where $\nu$ is a
characteristic rate of measurements.
With this definition, Eq. \r{1.20} holds both for ideal and nonideal
(realistic) measurements.
Relation \r{1.20} comes about since measurements (projections) 
dephase level 
$|e\rangle$, analogously to phase randomization by collisions, which
induce a linewidth that is equal to the collision rate $\nu$.

\begin{figure}[htb]
\centerline{\epsfig{file=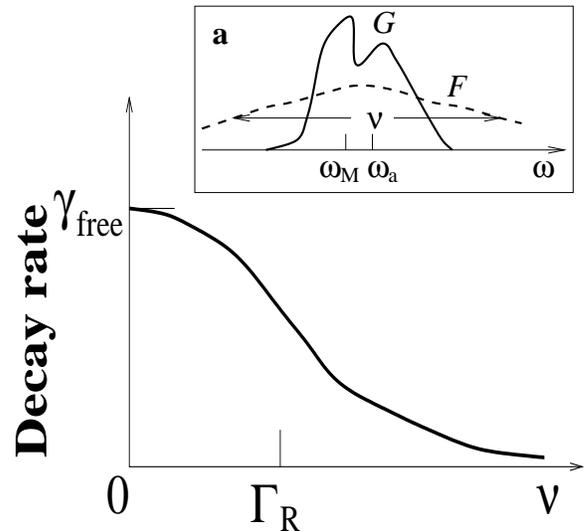,width=3.in}}
\protect\caption{
Decay rate as a function of the measurement rate $\nu$, under the QZE
conditions, shown in the inset [see Eq. \protect\r{26}].
}
\label{f4}
\end{figure}

A simple graphical analysis of the universal result Eq. \r{1.16} 
yields the main conclusions:

a) The QZE scaling (i.e., a {\em decrease of the decay rate $R$ with 
an increase of} $\nu$), is generally obtained when the measurement 
(dephasing) rate $\nu$ is much larger than the reservoir spectral 
width (Fig. \ref{f4}):
\be
\nu\gg\Gamma_R,|\omega_{\rm a}-\omega_{\rm M}|.
\e{11} 
Here $\Gamma_R$ is the reservoir width and $\omega_{\rm M}$ is the 
center of gravity of $G(\omega)$. 
In the special case of a peak-shaped $G(\omega)$, in Eq. \r{11}
$\omega_{\rm M}$ can be replaced by the position $\omega_{\rm m}$ of 
the maximum.
In the limit \r{11}, one can approximate the spectrum $G(\omega)$ by 
a $\delta$-function, with a constant $C$ being the integrated 
spectrum,
\be 
C=\int G(\omega)d\omega=\langle V^2\rangle.
\e{20}
This approximation becomes {\em exact} in the case of resonant Rabi 
oscillations ($\Gamma_R=0$, $\omega_{\rm a}=\omega_{\rm M}$), which 
explains why {\em the QZE is observable in Itano's experiment} 
\cite{ita90} {\em for any} $\nu$.
More generally, this approximation holds for any $G(\omega)$ that 
falls off faster than $1/\omega$ on the wings.
Then our universal expression yields the most general result for the 
QZE, namely that $R$ decreases with $\nu$:
\be
R\approx 2C/\nu,
\e{12}
where we defined generally
\be
\nu=[\pi F(\omega_{\rm a})]^{-1}
\e{30}
In particular, as follows from Eq. \r{1.19}, $\nu=2/\tau$ for 
instantaneous projections. 
The {\em flattening of the spectral peak} of $G(\omega)$ by the broad 
function $F(\omega)$ in the convolution is seen to be the origin of 
the QZE.
To put it simply, if the system is probed frequently enough, the QZE 
arises since {\em the effective decay
rate is averaged over all decay channels}, many of which are weak, due
to the energy uncertainty incurred by the measurements.

This result contradicts the claim of QZE universality and demonstrates
the failure of the standard quadratic expansion:
Eq. (8) shows that the QZE conditions can be much more stringent than
the requirement to have $t\sim 1/\nu\ll\tau_{\rm Z}$.
The crucial point emphasized below is that Eq. \r{11} may be {\em
principally impossible} to satisfy.

The above discussion presumes that the integral in \r{20} converges.
However, when $G(\omega)$ falls off as $1/|\omega-\omega_{\rm M}|$ or 
slower and hence $C$ is infinite, the QZE still holds under 
condition \r{11}, except that $R$ decreases with $\nu$ 
more slowly than $\nu^{-1}$.
This situation is illustrated by a peak-shaped reservoir with a 
slowly decreasing tail,
\be
G(\omega)=A|\omega-\omega_{\rm c}|^{-\beta}\ \ \text{for}\ \ 
s(\omega-\omega_{\rm c})\gg\Gamma_R
\e{21}
and $G(\omega)$ is cut off or diminishes fast for 
$s(\omega-\omega_{\rm c})<\Gamma_R$.
Here $\omega_{\rm c}$ is the cutoff frequency, $0<\beta<1$, 
$\theta()$ is the unit step function, and $s$ can equal 1 or $-1$.
For instance, Eq. \r{21} approximately holds near the waveguide 
cutoff ($s=1$, $\beta=1/2$) or the vibrational Debye cutoff 
($s=-1$, $\beta=1/2$).
The QZE scaling for a reservoir response \r{21} is found to be 
\be
R=B\nu^{-\beta}.
\e{22}
Here $B=[2^\beta\pi/\cos(\pi\beta/2)\Gamma(2+\beta)]A$, where
$\Gamma()$ is the gamma-function \cite{abr64} 
(in particular, $B=(8\pi^{1/2}/3)A$ for $\beta=1/2$).
Equation \r{22} holds for a sufficiently weak coupling
($A\ll\nu^{\beta+1}$) and under condition \r{11}, where, in the case
of \r{21}, $\omega_{\rm M}$ is replaced by $\omega_{\rm m}$ or,
equivalently, $\omega_{\rm c}$.

More generally, under condition \r{11}, the QZE scaling of $R$ occurs 
for any $G(\omega)$ such that 
\be
G(\omega)\rightarrow 0\ \ \mbox{at}\ \ \omega\rightarrow\infty.
\e{23}

Conditions \r{11} and \r{23} ensure the QZE scaling only
for sufficiently large measurement rates and {\em do not always
imply a monotonous decrease} of $R$ as $\nu$ increases (see Fig. 
\ref{f4}).
The latter behavior, which is what one usually has in mind in 
discussions of the QZE, is obtained only in special situations.
For instance, it occurs when $G(\omega)$ is a peak and 
$\omega_{\rm a}$ is within its width $\Gamma_R$, i.e., 
\be
|\omega_{\rm a}-\omega_{\rm M}|\alt\Gamma_R.
\e{26}

\begin{figure}[htb]
\vspace*{-1cm}
\centerline{\epsfig{file=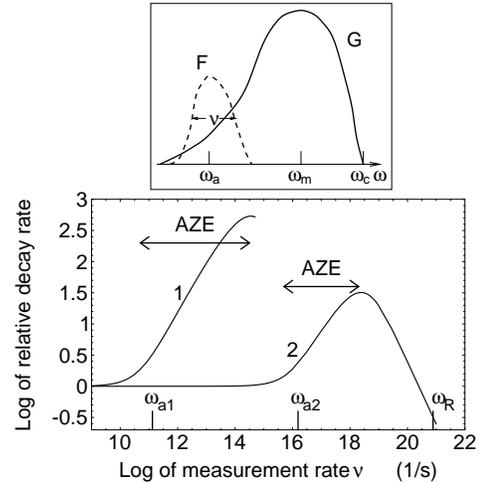,width=3.in}}
\vspace{-3cm}
\protect\caption{
AZE dependence of the decay rate.
Inset: Conditions for the AZE [Eq. \protect\r{13}]
Graph: the dependence of the logarithm of the normalized decay rate 
log$_{10}(R/R_{\rm RG}$) on log$_{10}\nu$ for a spontaneously
emitting hydrogenic state.
The values of the 
atomic transition frequencies corresponding to curves 1, 2: 
$\omega_{\rm a1}=1.32\times 10^{11}$ s$^{-1}$,
$\omega_{\rm a2}=1.55\times 10^{16}$ s$^{-1}$, whereas the
relativistic cutoff $\omega_{\rm R}=7.76\times 10^{20}$ s$^{-1}$. 
The corresponding Bohr frequencies are
$\omega_{\rm B1}=1.22\times 10^{15}$ s$^{-1}$, 
$\omega_{\rm B2}=8.50\times 10^{18}$ s$^{-1}$. 
The AZE ranges are marked.
}
\label{f5}
\end{figure}

b) The opposite to the QZE scaling is obtained 
whenever $\omega_{\rm a}$ is significantly detuned from 
the nearest maximum of $G(\omega)$ at $\omega_{\rm m}$, so that 
$G(\omega_{\rm a})\ll G(\omega_{\rm m})$. 
In the limit (Fig. \ref{f5} - inset):
\be
\nu\ll|\omega_{\rm m}-\omega_{\rm a}|.
\e{13}
the rate $R$ {\em grows} with $\nu$, since the 
dephasing function $F(\omega)$ is then probing more of the 
{\em rising} part of $G(\omega)$ in the convolution.
This limit implies the {\em anti-Zeno effect} (AZE) 
of decay acceleration by frequent measurements.
Physically, this means
that, as the energy uncertainty increases with the measurement rate
$\nu$, {\em the state decays into more and more channels, whose 
weight $G(\omega)$ is progressively larger}. 

Remarkably, we may impose condition (10) in {\em any reservoir} that
is not spectrally flat.
This reveals the {\em universality} of the AZE, which we noted already
for radiative decay in cavities \cite{kof96}.

As an example, consider the coupling spectrum of the form 
\bea
&&G(\omega)=A\omega^\eta,\ \ 0<\omega<\omega_{\rm C},\nonumber\\
&&G(\omega)=0\ \ \text{otherwise}, 
\ea{15}
so that $\omega_{\rm a}\ll\omega_{\rm C}$.
Using the Lorentzian $F(\omega)$ \r{3}, as obtained for realistic
(continuous) measurements (see Sec. \ref{III}), one obtains from Eq. 
\r{1.16} that for $\nu\ll\omega_{\rm C}$
\bem{34}
\be
R=\frac{2\pi A(\omega_{\rm a}^2+\nu^2)^{\eta/2}\sin\eta\chi}
{\sin\eta\pi}\ \ (0<\eta<1),
\e{34a}
\be
R=A\left[\nu\ln\frac{\omega_{\rm C}^2}{\omega_{\rm a}^2+\nu^2}+
\omega_{\rm a}\left(\pi+2\tan^{-1}\frac{\omega_{\rm a}}{\nu}\right)
\right]\ \ (\eta=1)
\e{34b}
\be
R=2\pi A\omega_{\rm a}^\eta+\frac{2A}{\eta-1}\nu
\omega_{\rm C}^{\eta-1}\ \ (\eta>1),
\e{34c}
\eml
where $\tan\chi=-\nu/\omega_{\rm a}$ ($0<\chi<\pi$).
Obviously, expressions \r{34} increase with $\nu$.

c) In the limit $\nu\rightarrow 0$, Eq. \r{1.16} yields the 
Golden Rule result (i.e., the normal decay rate in the absence of
measurements)
\be
R_{\rm GR}=2\pi G(\omega_{\rm a}).
\e{24}
More exactly, 
\be
R\approx R_{\rm GR}\ \ (\nu\ll\delta_{\rm a}).
\e{35}
Here $\delta_{\rm a}$ is the interval around $\omega_{\rm a}$ over 
which $G(\omega)$ changes appreciably, when $G(\omega)$ increases not
faster than a linear function with the decrease of 
$|\omega_{\rm m}-\omega|$, and may be much less otherwise.
In particular, as follows from Eqs. \r{34}, in the case \r{15}
$\delta_{\rm a}=\omega_{\rm a}$ if $\eta\le 1$ and 
$\delta_{\rm a}=\omega_{\rm a}^\eta/\omega_{\rm C}^{\eta-1}\ll
\omega_{\rm a}$ if $\eta>1$.
 
d) More subtle behavior occurs in the domain between the QZE and AZE
limits.
Assume, for simplicity, that $G(\omega)$ is 
single-peaked and satisfies condition \r{23}.
When $\nu$ increases from the limit \r{13} up to the range where the
right-hand-side inequality is violated, then
$\nu\gg|\omega_{\rm m}-\omega_{\rm a}|$, which is now equivalent to
condition \r{11}, implying the Zeno scaling of Eq. \r{12} or \r{22}. 
But even in this QZE-scaling regime, $R$ {\em remains larger than the 
Golden Rule rate} $R_{\rm GR}$ \r{24}, up to much higher $\nu$, as
expressed by the following condition for {\em ``genuine QZE''}
\be
R<R_{\rm GR}\ \ \text{for}\ \ \nu>\nu_{\rm QZE},
\e{27}
where
\bem{25}
\be
\nu_{\rm QZE}=2C/R_{\rm GR}=C/\pi G(\omega_{\rm a})
\e{25a}
in the case of a finite $C$, and
\be
\nu_{\rm QZE}=(B/R_{\rm GR})^{1/\beta}=[B/2\pi 
G(\omega_{\rm a})]^{1/\beta}
\e{25b}
\eml
in the case of Eq. \r{21}.
The quantity $\nu_{\rm QZE}$ may turn out to be much greater than the 
boundary $\nu_1$ of the QZE-scaling regime (see below).

The value of $\nu_{\rm QZE}$ given by Eq. \r{25a} was identified with 
the reciprocal ``jump time'', i.e., the maximal time
interval between measurements for which the decay rate is appreciably 
changed \cite{sch97}.
However, the {\em correct value} of the reciprocal jump time is rather
$\delta_a$, which may be smaller by many orders of magnitude than 
$\nu_{\rm QZE}$.
In the special case of ideal instantaneous measurements and a 
Lorentzian or Lorentzian-like $G(\omega)$, the genuine-QZE
condition \r{27} reduces to that of Ref. \cite{fac}.

These considerations apply (with some limitations) to 
hydrogenic radiative decay (spontaneous emission), for which 
$G(\omega)$ can be calculated {\em exactly} \cite{mos72}:
\be
G(\omega)=\frac{\alpha\omega}{[1+(\omega/\omega_{\rm B})^2]^4}.
\e{18}
Here $\omega_{\rm B}\sim c/a_{\rm B}$, where $c$ is the vacuum light
speed and $a_{\rm B}$ is the radius of the electron orbit.
Then Eqs. \r{1.16} and \r{1.19} may yield the AZE trend 
\be
R\approx\alpha\nu[\ln(\omega_{\rm B}/\nu)+C_1]\ \ 
(\omega_{\rm a}\ll\nu\ll\omega_{\rm B}),
\e{1}
where $C_1=0.354$ and $\nu=2/\tau$.
The AZE trend should be observable (Fig. \ref{f5}) for 
$\nu\agt\omega_{\rm a}$, i.e., for microwave 
Rydberg transitions on a ps scale (provided we can isolate one
transition) and for optical transitions on the sub-fs scale.
The boundary between the AZE and QZE-scaling regions is now given by 
$\nu_1\sim\omega_{\rm B}$ and the genuine-QZE condition \r{27} by
$\nu>\nu_{\rm QZE}\sim\omega_{\rm B}^2/12\pi\omega_a\gg\nu_1$ [cf. 
Eq. \r{25a}], rendering $R<R_{\rm GR}=2\pi\alpha\omega_a$.

This analysis implies that the ``genuine QZE'' range 
$\nu>\nu_{\rm QZE}$ is {\em principally unattainable}, since it 
requires measurement rates above the relativistic cutoff 
$\omega_{\rm R}$, which are {\em detrimental} to the system, 
leading to the production of new particles.
A similar principal obstacle occurs for radioactive decay.
By contrast, the AZE is {\em accessible} in decay processes, such as
spontaneous emission or 
the nuclear $\beta$-decay, and can essentially always be imposed.

\section{Realistic measurements}
\label{III}

\begin{figure}[htb]
\centerline{\epsfig{file=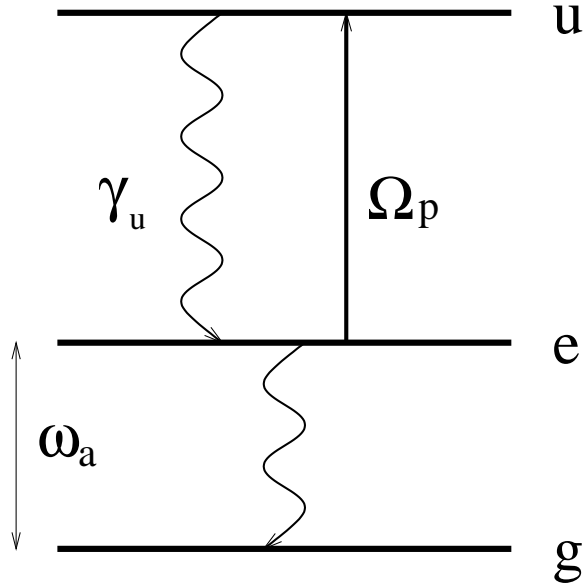,width=3.in}}
\protect\caption{
The Cook scheme: see text for explanations.
}
\label{f6}
\end{figure}

Thus far we have assumed ideal instantaneous projections on 
$|e\rangle$.
Does a more
realistic description of measurements still support these results?
The answer is {\em positive} for the two possible types of
measurements of $|e\rangle$: 

a) {\em Impulsive measurements}. 
Such measurements are realizable, e.g., in Cook's scheme \cite{coo88}
implemented by Itano et al. \cite{ita90} (Fig. \ref{f6}): the decay 
process is repeatedly interrupted by a short pulse 
transferring the population of $|e\rangle$ to a higher auxiliary state
$|u\rangle$, which then decays back fast enough to $|e\rangle$ {\em 
incoherently}.
This case conforms to our universal result Eq. \r{1.16} to a very 
good approximation, with the dephasing function given by Eq. \r{1.19}.

b) {\em Continuous measurements}.
These measurements should {\em not be confused with} 
the limit of vanishing intervals between successive continuous 
projections discussed by Misra and Sudarshan.
This limit is {\em unphysical},
corresponding to an infinite energy spread $\hbar\nu$.
In contrast, realistic continuous measurements, though monitoring the
state incessantly, still require {\em a finite time} for completing
an observation, i.e., they have a {\em finite effective rate} $\nu$.

\begin{figure}[htb]
\centerline{\epsfig{file=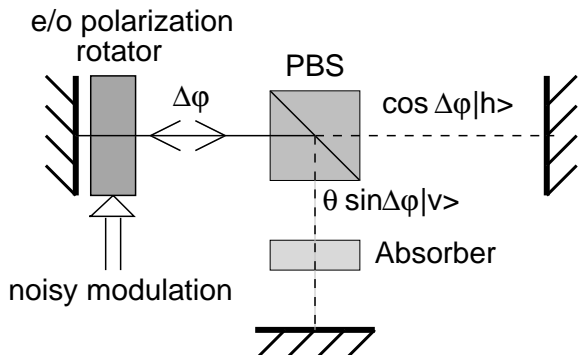,width=3in}}
\protect\caption{
Setup for controlling the polarization decay of a
photon bouncing between the mirrors. Measurements are effected by a
polarization beam-splitter (PBS) and an absorber with variable
transparency $\theta$. 
The ``reservoir'' into which the polarization decays is realized upon 
modulating a Pockels cell (which rotates the polarization by $\df$) 
by a field with controllable noise properties.
}
\label{f7}
\end{figure}

This is seen for {\em stationary dephasing} 
when the $|e\rangle\leftrightarrow|u\rangle$ transition is driven by 
a near-resonant continuous-wave field, such that 
$1/\Omega_p\gg 1/\gamma_u$, yielding
\be
\nu=2\Omega^2/\gamma_{u}.
\e{2}
One can show that the universal result still holds for such 
measurements, with 
$F(\omega)$ being a Lorentzian centered at $\omega_{\rm a}$:
\be
F(\omega)=\frac{1}{\pi}\frac{\nu}{(\omega-\omega_a)^2+\nu^2}.
\e{3}

Thus, qualitatively, there is no essential distinction between 
different frequent measurements: ideal projections, impulsive 
measurements, and continuous measurements. 

\section{QZE and AZE for Photon Polarization Dephasing}

As an example, consider the decay of photon polarization into a 
``reservoir'' created by the phase-randomizing element (polarization
rotator) controlled by a noisy field (see Fig. \ref{f7}) \cite{kwi98}.

Let us first assume constant phase jumps $\Delta\varphi$ at the
polarization rotator.
Uninterrupted evolution then corresponds to Rabi oscillations of the
horizontal polarization probability (for $\theta=1$):
\be
P_{h}(n) = \cos ^{2}(n\Delta \varphi).
\e{4}
Decay via perfect (projective) measurements (for $\theta=0$) results
in
\be
P_{h}(n) = e^{-n(\Delta \varphi)^2}.
\e{expdecay}
This decay is slower than Rabi oscillations, signifying the QZE.
Imperfect (weak) measurements ($0<\theta<1$) yield for small phase
jumps [$(\Delta\varphi)^2\ll(1-\theta)^2$] an exponential decay
\be
P_{h}(t=n\tau_{\rm r})=\exp\left[-\frac{2(\Delta\varphi)^2}
{\tau_{\rm r}^2\nu}t\right],
\e{5}
$\tau_{\rm r}$ being the round-trip time.
This decay still conforms with the QZE: the decay rate decreases with
the measurement rate
\be
\nu=\frac{2(1-\theta)}{1+\theta}\frac{1}{\tau_{\rm r}}.
\e{6}

We now proceed to discuss random phase jumps $\Delta\varphi_n$, caused
by noisy modulation.
We then obtain for sufficiently small jumps
\be
R=2\pi\int_{-\pi/\tau_{\rm r}}^{\pi/\tau_{\rm r}}d\omega
G(\omega)F(\omega),
\e{7}
which is an extension of Eq. \r{1.16} to such a ``reservoir''.
Here $F(\omega)$ is peaked at $\omega=0$ and has the characteristic
width \r{6}.

The key parameter for $G(\omega)$ in Eq. \r{7} is the correlation of 
consecutive phase jumps:
\be
\langle\Delta\varphi_{n+1}\Delta\varphi_n\rangle=
\gamma\langle\Delta\varphi_n^2\rangle.
\e{8}
For highly correlated jumps $(\gamma\approx 1)$ we find
\be
G(\omega)\approx\frac{B^2}{\pi\tau_{\rm r}^2}\frac{\Gamma_{\rm R}}{
\Gamma_{\rm R}^2+\omega^2},
\e{9}
i.e. a Lorentzian of width $\Gamma_{\rm R}=(1-\gamma)/\tau_{\rm r}$.
For highly anticorrelated jumps $(\gamma\approx -1)$ the ``reservoir''
spectrum is a sum of two shifted Lorentzians,
\be
G(\omega)\approx\sum_{j=\pm 1}\frac{B^2}{\pi\tau_{\rm r}^2}
\frac{\Gamma_{\rm R}'}{
{\Gamma_{\rm R}'}^2+(\pi/\tau_{\rm r}+j\omega)^2}
\e{14}
of width $\Gamma_{\rm R}'=(1+\gamma)/\tau_{\rm r}$.

\begin{figure}[htb]
\vspace*{-3cm}
\centerline{\epsfig{file=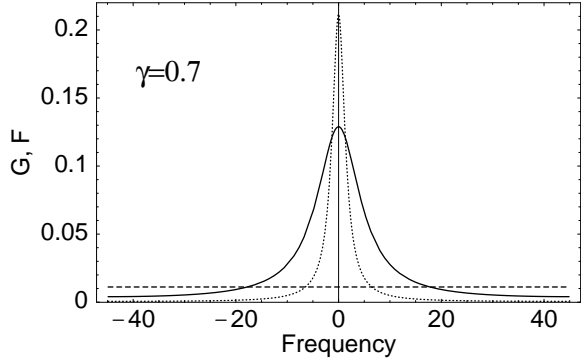,width=3.in}}
\vspace{-2.5cm}
\protect\caption{
The overlap of $G$ (solid line) and $F$ for correlated phase jumps in 
the setup of
Fig. \protect\ref{f7} [see Eqs. \protect\r{7}-\protect\r{16}].
Dashed line: $F(\omega)$ with $\theta=0$ (perfect projections). 
Dotted line: $F(\omega)$ with $\theta=0.9$ (ineffective 
measurements). 
Here $B=0.1,\ \tau_{\rm r}=0.07$. 
}
\label{f8}
\end{figure}

\begin{figure}[htb]
\vspace*{-3cm}
\centerline{\epsfig{file=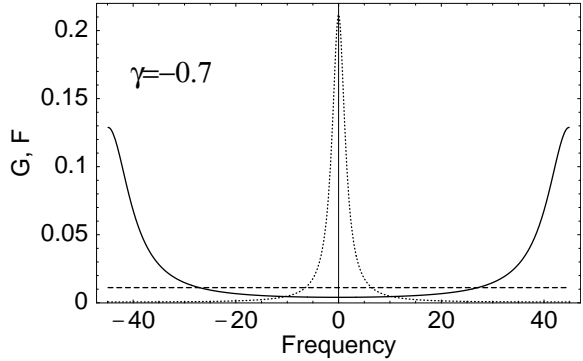,width=3.in}}
\vspace{-2.5cm}
\protect\caption{
Idem, for anticorrelated phase jumps.
}
\label{f9}
\end{figure}

If $G(\omega)$ is peaked at $\omega=0$ (the high-correlation case 
$\gamma\approx 1)$, then {\em perfect measurements} $(\theta=0)$,
corresponding to a flat $F(\omega)$, will {\em reduce} $R$, causing 
the QZE trend, as compared to weak measurements $(\theta\approx 1)$,
corresponding to a narrow spectral profile
\be
F(\omega)\approx\frac{1}{\pi}\frac{\nu}{\nu^2+\omega^2},
\e{16}
with $\nu=(1-\theta)/\tau_{\rm r}$ (see Fig. \ref{f8}).

\begin{figure}[htb]
\vspace*{-2.5cm}
\centerline{\epsfig{file=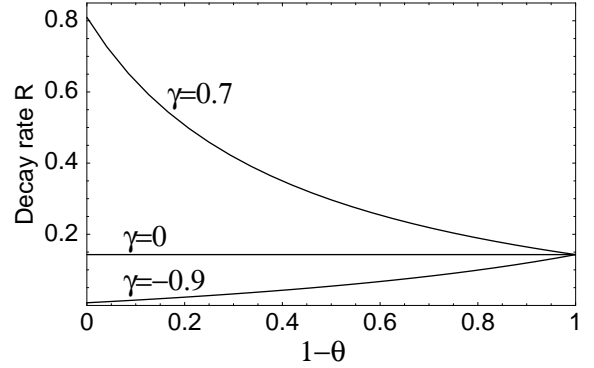,width=3.in}}
\vspace*{-2.5cm}
\protect\caption{
The decay rate in the setup of Fig. \protect\ref{f7} as a function of 
the measurement effectiveness
$1-\theta$ (which is on the order of the dimensionless measurement
rate $\nu\tau_{\rm r}$) for correlated ($\gamma=0.7$), Markovian 
($\gamma=0$) and anticorrelated ($\gamma=-0.9$) phase fluctuations.
}
\label{f10}
\end{figure}

The {\em opposite} (AZE) trend holds if $G(\omega)$ is peaked at 
$\omega=\pm\pi/\tau_{\rm r}$ (the highly anti-correlated case, 
$\gamma\approx -1)$ (Fig. \ref{f9}).
The two trends are illustrated in Fig. \ref{f10}.

\begin{figure}[htb]
\centerline{\epsfig{file=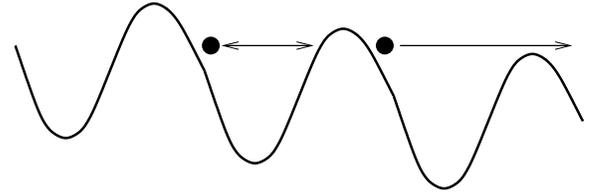,width=3.in}}
\protect\caption{
Atoms tunnel out of a ``washboard'' potential.
}
\label{f11}
\end{figure}

\begin{figure}[htb]
\vspace*{-3.cm}
\centerline{\epsfig{file=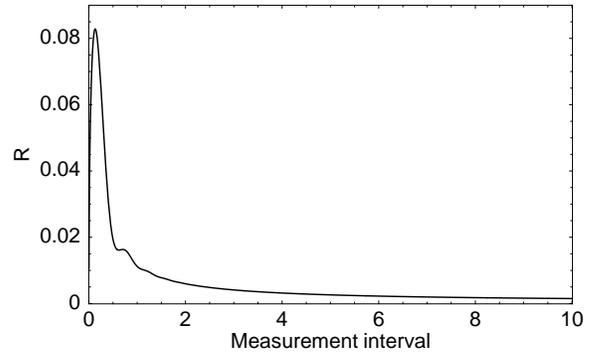,width=3.in}}
\vspace{-2.5cm}
\protect\caption{
The AZE trend for the decay rate $R$ (in units of $\tau_0^{-1}$) in a 
``washboard'' potential (Fig. 
\protect\ref{f11}) as a function of the scaled measurement
interval $\tau/\tau_0$ ($\omega_g=10/\tau_0$, 
$M^2a/8\hbar^2k_L^3=0.01$, where $M$ is the mass of the atom).
}
\label{f12}
\end{figure}

\onecolumn\twocolumn[
\widetext
\begin{table}
\caption{}

\begin{tabular}{l|l|c|c}
\bf Process&\bf References&$\tau_{\rm QZE}$&$\tau_{\rm AZE}$\\
\hline\hline
Radiative decay in a cavity&Kofman \& Kurizki \protect\cite{kof96}
&ns&ns\\
\hline
Radiative decay&Kofman \& Kurizki \protect\cite{kof00}&does not exist
&ps (Rydberg transition)\\
in open space&&&fs (optical transition)\\
\hline
Photon polarization&Kofman, Kurizki&10 ns&10 ns\\
decay via random&\& Opatrny \protect\cite{kwi98}&&\\
modulation in a cavity&&&\\
\hline
Transmission of tunneling&Japha \& Kurizki \protect\cite{jap96}&?&ns\\
emitting atoms&&&\\
\hline
Atomic tunneling&Wilkinson et al. \protect\cite{wil97}&0.01 $\mu$s
&$\mu$s\\
(escape) from&&&\\
``washboard'' (accelerated)&&&\\
periodic potential&&&\\
\hline
Electron tunneling in&Silvestrini et al. \protect\cite{sil97}&?&\\
current-biased SQUID&&&10 ns\\
(``washboard'' potential)&&&\\
\hline
Nuclear $\beta$-decay&Kofman \& Kurizki \protect\cite{kof00}&does not
exist&10$^{-18}$ s\\
\hline
Near-threshold&Lewenstein&fs&ms\\
photodetachment&\& Rz\c{a}\.{z}ewski \protect\cite{lew00}&&\\
\end{tabular}
\label{t1}\end{table}
]
\narrowtext

\begin{figure}[htb]
\vspace*{-3.cm}
\centerline{\epsfig{file=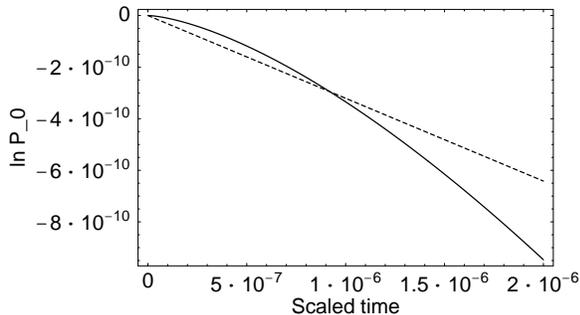,width=3.in}}
\vspace{-2.5cm}
\protect\caption{The decay law in a ``washboard'' potential as a
function of the scaled time $t/\tau_0$ 
($\omega_g=10/\tau_0$, $M^2a/8\hbar^2k_L^3=0.01$).
Dashed line: the exponential decay with the rate $R_{\rm GR}=3.2\times
10^{-4}\tau_0^{-1}$.
}
\label{f13}
\end{figure}

\section{Atom escape (decay) by tunneling from accelerated potential}
\label{V}

Theoretical and experimental studies by Raizen et al. \cite{wil97} 
have shown that cold atoms trapped in an accelerated 
(``washboard''-shaped) periodic potential tunnel out (Fig. \ref{f11}) 
non-exponentially, at times $t\alt\tau_0\equiv\omega_g/(k_La)$, where 
$\hbar\omega_g$ is the band gap which equals approximately half the
potential well depth, $a$ the acceleration and $k_L$ the 
wave number of the laser creating this potential.
Our prediction is that AZE should arise for measurement intervals 
$\tau\gg 1/\omega_g$ and QZE for $\tau\ll 1/\omega_g$.
These trends are plotted in Fig. \ref{f12}.
Figure \ref{f13} shows (in the logarithmic scale) the decay law 
$P_0(t)=|\alpha(t)|^2$ for short times (the solid curve).
The two curves in Fig. \ref{f13} intersect at the time 
$\tau_{\rm QZE}=2/\nu_{\rm QZE}$, so that the genuine QZE occurs for 
$\tau<\tau_{\rm QZE}$.
Note that in this case $\tau_{\rm QZE}$ ($\approx 10^{-6}\tau_0$) is 
much less than the boundary of the QZE-scaling region 
$1/\omega_g=0.1\tau_0$.

\section{Experimental accessibility}

Table \ref{t1} lists a number of processes where the AZE may be 
observed at accessible measurement intervals $t\sim\tau_{\rm AZE}$.
By contrast, the QZE ($\tau_{\rm QZE}\ll\tau_{\rm AZE}$) either is
principally unobservable or, as a rule, is practically much less
accessible than the AZE ($\tau_{\rm QZE}\ll\tau_{\rm AZE}$).
Notable exceptions, where the AZE and QZE are equally accessible, are 
the cases of a narrow (near-resonance) reservoir, such as radiative or
photon-polarization decay in a cavity.

\section{Conclusions}

Our simple universal formula \r{1.16} results in general criteria for 
the QZE:\\
(i) It can only occur in systems with spectral width below the 
resonance energy.\\
(ii) It is {\em principally unattainable} in open-space radiative or 
nuclear $\beta$-decay, because the required measurement rates would 
cause the creation of new particles.\\ 
(iii) Contrary to the widespread view, frequent measurements
can be chosen to {\em accelerate} essentially any decay process. 
Hence, the {\em anti-Zeno effect} should be {\em far more 
ubiquitous than the QZE.}

A variety of systems have been shown to be promising candidates for
experimental studies of the AZE, which is almost always much more
accessible than the QZE.

\end{document}